\documentstyle[version2,preprint,aps]{revtex}

\makeatother

\begin{document}

\draft
\preprint{\today}
\begin{title}
Dissipative Chaos in Semiconductor Superlattices
\end{title}
\author{ Kirill N. Alekseev$^{a-c}$, Gennady P. Berman$^{a-c}$, David K. Campbell$^c$ \\
Ethan H. Cannon$^c$, and Matthew C. Cargo$^c$}

\begin{instit}
$^a$Center for Nonlinear Studies and Theoretical Division, Los Alamos National
Laboratory \\
Los Alamos, New Mexico 87545, U.S.A.\\
$^b$Kirensky Institute of Physics, 660036, Krasnoyarsk, Russia \\
$^c$Department of Physics,
University of Illinois at Urbana-Champaign,\\
1110 West Green St.,
Urbana, IL 61801-3080, U.S.A.
\end{instit}
\receipt{April 24, 1995; Revised \today}

\begin{abstract}

We consider the motion of ballistic electrons in a miniband of a semiconductor
superlattice (SSL) under the influence of an external, time-periodic electric field.
We use the semi-classical balance-equation approach which incorporates
elastic and inelastic scattering (as dissipation) and the self-consistent field
generated by the electron motion. The coupling of electrons in the miniband to
the self-consistent field produces a cooperative nonlinear oscillatory mode which,
when interacting with the oscillatory external field
and the intrinsic Bloch-type oscillatory mode, can lead to complicated dynamics,
including
dissipative chaos. For a range of values of the dissipation parameters
we determine the regions in the amplitude-frequency plane of the external
field in which chaos can occur. Our results suggest that for terahertz
external fields of the amplitudes achieved by present-day free electron lasers,
chaos may be observable in SSLs. We clarify the nature of this novel nonlinear dynamics 
in the superlattice-external field system by exploring analogies to the
Dicke model of an ensemble of two-level atoms coupled with a resonant
cavity field and to Josephson junctions.
\end{abstract}

\renewcommand{\baselinestretch}{1.656}   


\narrowtext

\section{INTRODUCTION}

More than two decades ago, Esaki and Tsu \cite{esaki} discovered
a striking nonlinear effect in semiconductor superlattices (SSL's),
establishing that the dissipative motion of electrons within a single SSL
miniband in the presence of a {\it static} electric field can produce a
{\it negative differential
conductivity} (NDC) in the stationary current-voltage characteristic.
In view of its potential applications to ultra-small
electronic devices \cite{ferry1}, interest in this
effect remains high \cite{balkan}, and it continues to be studied by
a variety of different theoretical and experimental techniques
\cite{sibille1,sibille2}.
The importance of these systems is reflected, for
example, in the many articles in semi-popular physics
literature \cite{capasso}.

In the past several years, technological advances in semiconductor nanostructure
fabrication and in electromagnetic field generation techniques have made possible
detailed studies of a wide range of other nonlinear phenomena involving electron
transport in SSLs
in the presence of a variety of electromagnetic (EM) fields \cite{sibille1,sibille2}.
Very recently, the effect of alternating fields
in the terahertz (THz) domain on the nonlinear current-voltage
characteristics of superlattices has been investigated experimentally
\cite{ignatov1} and observations of photon-assisted resonant tunneling and negative
{\it absolute} resistance have been reported \cite{guimaraes}.

These extensive experimental efforts have been paralleled by many related theoretical
studies \cite{epstein1,vatova,epstein2,epstein3,bass1,ignatov2,holthaus,bass2,bass3}.
These studies have examined the propagation of the electromagnetic
solitons through the superlattice \cite{epstein1}; self-consistent nonlinear
plasma oscillations \cite{vatova,epstein2,epstein3};
Hamiltonian chaotic
dynamics of electrons in a constant magnetic field interacting with
electromagnetic waves \cite{bass1}; emission of electromagnetic
radiation from superlattices due to multiphoton transitions \cite{ignatov2}.
Although most of these studies have been in the semi-classical
regime, the consequences of a fully quantum mechanical treatment
of miniband transport
in an intense oscillating EM field have also been examined \cite{holthaus}.
Reviews of these and other theoretical studies are given in \cite{bass2,bass3}.
All these investigations have dealt with the {\it intra}-miniband dynamics
of the electron, neglecting the possibility of {\it inter}-miniband transitions;
additional nonlinear phenomena can become relevant if inter-miniband
dynamics are considered.

In a previous paper \cite{alekseev1}, we extended earlier studies
of the {\it intra}-miniband dynamics of
ballistic electrons in an SSL with an ac external electric field by taking
into account the self-consistent EM field generated by the electron current.
The coupling of the electron motion to the
self-consistent field results in cooperative oscillations
which alter the dynamics substantially. In particular, we showed
that, with the inclusion of the self-consistent field, the inherently nonlinear
nature of the electron dynamics implies that, under the influence of the
external ac field, various dynamical instabilities,
including transitions to chaos, can occur \cite{alekseev1}.
However, our previous work was limited to the {\it non-dissipative} regime
and thus was strictly valid only on time-scales shorter than all
characteristic relaxation times for the electron's energy and momentum.
Hence, {\it a priori} we could not determine the long-time (stationary) dynamical
behavior. More importantly, the assumption of no dissipation
is of dubious applicability to
the interpretation of experiments in real systems, in which dissipative 
effects are certainly present at some level. 

In the present study, we generalize our previous work by considering
explicitly the role of dissipation in the balance equations
describing the SSL plus field system. After recalling briefly
how the conventional balance equations recover the standard results of
Bloch oscillations and negative differential conductivity, we
show that our extension of the balance equations to
account for collective effects on the electron's 
motion can (for certain conditions) result in chaotic dynamics,
both transient and stationary, even in the presence of
damping. We discuss and exploit analogies between
the electrons in an SSL plus field system and the dynamics of lasers and
optical bistable systems and of damped, driven Josephson
junctions; the extensive tradition of studying
dissipative nonlinear dynamics in these problems
allows us to use these analogies to guide our studies
of the semiconductor system.

The remainder of the paper is organized into three sections.
In Section 2 we use a balance equation approach to
formulate (phenomenological) equations of the motion for the
electron in the SSL in the presence of an external ac field,
including the self-consistent field generated by the electron current.
We show how to incorporate various dissipative channels
in these equations and illustrate that the solutions to these
equations capture the anticipated behavior in several simple
limiting cases, including those of (undamped) Bloch oscillations
and (stationary) negative differential conductivity. In preparation
for our study of the general case, we introduce a re-scaled
(pseudo-spin) representation of the equations of motion and
demonstrate the analogy with the optical systems, in which
chaotic behavior is well established. 
In Section 3 we present the details of our study of the dissipative
chaotic dynamics of the SSL plus field system. First,
we show that our equations reduce
(in two different limits) to the equations of two well-known chaotic
systems, (1) the
Lorenz equations and (2) the damped, driven pendulum. Second, we
present numerical studies of the full equations. Our results
suggest that, for a wide range of values of the frequency and
amplitude of the external field and for experimentally
relevant values of the damping, stationary chaotic
dynamics should be observed, interspersed with transient
chaos leading eventually to periodic motion. We provide
detailed confirmation of the chaotic  behavior in the form
of Lyapunov exponents, power spectra, and direct time evolution
traces. In Section 4 we summarize our results,
examine critically the possible parameter ranges in which the effects
we predict can be observed in experiments, discuss how these
effects might manifest themselves in experimental
observables, and mention a number of
open theoretical issues.

\section{BALANCE EQUATION APPROACH}

Consider the motion of ballistic electrons in an SSL in the presence
of an external ac electric field with amplitude $E_0$ and
frequency $\Omega$,
$$
E_{ext}(t)=E_0\cos\Omega t,\eqno(1)
$$
which is applied in the direction perpendicular to the layers of the SSL,
{\it i.e., } in the direction of the motion within the SSL miniband.
In the standard tight-binding approximation,
the dispersion relation of the electron belonging to a single miniband in the
SSL is \cite{esaki,bass2}

$$
\varepsilon(p)={{\Delta}\over{2}}\left[1-\cos\left({{pa}\over{\hbar}}\right)\right],\eqno(2)
$$
where $\Delta$ is the miniband width, and $a$ is the SSL periodicity.
This non-quadratic dependence of the electron energy on the
quasi-momentum within the miniband renders the dynamics of electrons inherently
nonlinear.

Generalizing the balance equation approach of Ref. \cite{ignatov2}
to incorporate the self-consistent field \cite{scholl},\cite{ferry2}
generated by the electron current,
we find that (with the assumption of spatial homogeneity)
the equations describing
the electron's motion in the combined external and self-consistent field become
$$
\dot V=-eE_{tot}(t)/m(\varepsilon)-\gamma_vV,\eqno(3a)
$$
$$
\dot\varepsilon=-eE_{tot}(t)V-\gamma_\varepsilon(\varepsilon-\varepsilon_0),\eqno(3b)
$$
$$
\dot E_{sc}=-4\pi j-\alpha E_{sc},\eqno(3c)
$$
\noindent
where
$$
E_{tot}(t)=E_{sc}(t)+E_{ext}(t) \quad {\rm and } \quad j \equiv -eNV.\eqno(3d)
$$
In Eqs. (3), $V$ is the average velocity (along the SSL axis)
and $N$ is the number of
electrons per unit volume, so that $j$ is the average
electron current density; $\varepsilon$ is the average
energy of the electrons; $\varepsilon_0$ is the equilibrium
energy of carriers (resulting from thermal energy and/or external pumping);
and $E_{sc}$ is
the self-consistent EM field generated by the electron current $j$.
All quantities are measured in $CGS$ units.

Let us say a few words about the physical interpretation of
these equations. Eq. (3a) is the basic equation of motion of electrons
belonging to one
miniband (with the dispersion law (2)) in the presence
of the electric field $E_{tot}(t)$.
Importantly, the dependence of the electron's effective
mass, $m(\varepsilon)$, on energy in (3a) is \cite{ignatov2}
$$
m(\varepsilon)={{m_0}\over{1-2\varepsilon/\Delta}},\quad m_0\equiv{{2\hbar^2}\over{\Delta a^2}},\eqno(4)
$$
where $m_0$ is the electron's effective mass at the bottom of the
miniband. Note that $m(\varepsilon)$ may take negative values
for $\varepsilon>\Delta/2$. The occurrence of a negative effective mass is
connected with the existence of negative differential drift velocities
(see, {\it e.g.}, \cite{ignatov2}
and references therein).

Eq. (3b) describes the heating ({\it i.e.},
increase in energy) of the
electrons by the field $E_{tot}(t)$ and cooling ({\it i.e.,} relaxation of
energy)
to the equilibrium value $\varepsilon_0$.
The thermal equilibrium energy value, $\varepsilon_0$, has the
temperature dependence \cite{ignatov2,suris}
$$
\varepsilon^{(T)}_0={{\Delta}\over{2}}\left[1-{{I_1(\Delta/2k_BT)}\over{I_0(\Delta/2k_BT)}}\right],\eqno(5)
$$
where $I_{0,1}$  are the modified Bessel functions, $T$ is lattice temperature,
and $k_B$ is Boltzmann's constant.

The parameters $\gamma_v$, $\gamma_{\varepsilon}$ and $\alpha$ are the
{\it phenomenological} relaxation constants
of the average velocity, energy and the self-consistent field, respectively.
The treatment of these relaxation parameters as constants,
independent of energy, is an approximation; we shall discuss our present
understanding of the validity of this approximation in the concluding
section. Physically, in an SSL, the damping of the average velocity
is mainly due to the elastic scattering of
electrons with the impurities, structural disorder, and interface roughness.
Typically, the main channel of energy dissipation for the
electron sub-system is
inelastic phonon scattering.

Eq. (3c) is Maxwell's equation for the
time evolution of the self-consistent electric field, with
an additional term describing its relaxation.
The form of this
equation, including the phenomenological relaxation
constant, $\alpha$ -- which models effects of interactions of the
self-consistent field with degrees of freedom beyond the ballistic
electrons described dynamically in our equations -- is familiar
from the literature on bulk semiconductors \cite{scholl}.
Examples of the processes contributing to $\alpha$ include
surface effects and the generation of polar phonons due to
the finite polarization of the crystal produced by the field.

Following the original deviations of the ``balance equations'' (3a) and (3b)
in \cite{ignatov3,lei1},
several articles have discussed the {\it microscopic}
derivation and justification
of the equations and their use in modeling both of general
transport properties in SSLs \cite{lei2} and of specific
nanoscale devices \cite{ferry1}, \cite{ferry3}.
For our present purposes, however, it is sufficient
to consider Eqs. (3) as a phenomenological set of equations.

Before analyzing the nonlinear dynamics of the full system (3) in detail,
let us examine briefly several important limiting cases, in order
to illustrate the consistency of the balance equation
approach with known results. For simplicity, we shall momentarily
ignore the self-consistent field equation entirely and
focus on the consequences of Eqs. (3a) and (3b) alone.
Consider first the case in which
the electrons in the SSL are influenced only by a {\it constant}
external field
$E_{tot}=E_{ext}=E_0=const$, and for
which the relaxation processes can be neglected: {\it i.e.},
$\gamma_{\varepsilon} = \gamma_v = 0$. In this simple limiting case,
a straight-forward calculation shows that the electrons perform
harmonic oscillations with velocity $V=V_0\sin\omega_st$. These
are the familiar ``Bloch oscillations'' \cite{esaki,capasso},
and the characteristic frequency of these oscillations $\omega_s=eaE_0/\hbar$
is known as the Bloch \cite{ignatov1} or Stark \cite{bass2}
frequency. For typical SSL's
and for typical electric fields, $(\sim 1 - 10 kV/cm)$,
the Bloch frequency belongs to the THz domain.
Although there remains some controversy, experimental evidence
for these Bloch oscillations has recently been reported \cite{feldman}.

Consider next the case of in which there is still a constant electric field and
the relaxation effects are also included, so that
$\gamma_v\not= 0$, $\gamma_\varepsilon\not= 0$. This problem
was first considered by Esaki and Tsu \cite{esaki} for
the particular case $\gamma_v=\gamma_\varepsilon$, and the generalization
to the case $\gamma_v\not =\gamma_\varepsilon$ is given
in \cite{ignatov2,ignatov3,lei1}. Again, a straightforward
calculation shows that the system undergoes damped oscillations
and that in the steady-state the current $ j = -eNV$ becomes
$$
j=\frac{e^2NE_0/\gamma_v}{1+((eE_0a/\hbar)^2/(\gamma_v \gamma_{\varepsilon}))}\frac{I_1(\Delta/2k_BT)}{I_0(\Delta/2k_BT)} \frac{1}{m_0},\eqno(6)
$$
which reduces to the original Esaki-Tsu result \cite{esaki}
in the zero temperature limit (so the ratio of Bessel functions goes
to one) and when we choose $\gamma_v= \gamma_{\varepsilon} \equiv 1/\tau$.
In this limit, it is easy to see that
$\partial j/\partial E_0<0$ for $(eaE_0\tau/\hbar)>1$, so
that there NDC in this regime.
Experimentally, both the NDC effect in SSL \cite{sibille1}
and the effect of thermal saturation of miniband transport in an
SSL \cite{brozak},
due to the dependence of the equilibrium electron energy
$\varepsilon^{(T)}_0$ ({\it cf.} Eq. (5))
on temperature have been observed.
In this regard, it
is interesting to note that the generalization of the formula of Esaki and Tsu
to finite temperature and to $\gamma_v \neq \gamma_{\varepsilon}$
describes with reasonable accuracy the stationary transport properties of
an SSL, even in the case when the energy of the external field and the thermal
energy of the electrons are comparable to the miniband width and the
quasiclassical description becomes {\it a priori} inadequate \cite{sibille2}.

We have considered these two special cases in the absence of the
self-consistent field, {\it i.e.}, $E_{sc}(t)\equiv 0$. Importantly,
if we include  $E_{sc}(t)$, we again find Bloch oscillations (albeit
{\it not} simple sinusoidal motion) in the case of no relaxation effects
and NDC for the steady state current-voltage characteristic when
relaxation effects are included. We do not present these results in detail
here \cite{cannon}, as they are not essential to our present study.

To prepare for our study of the time-varying external field case,
including  $E_{sc}(t)$, we introduce some scalings of the
variables in Eq. (3), for these will both simplify the analysis and make
apparent an important analogy with the nonlinear dynamics in optical systems.

We introduce new variables
$$
v={{2\hbar}\over{\Delta a}}V,\quad w={{\varepsilon-\Delta/2}\over{\Delta/2}}, \quad w_0={{\varepsilon_0-\Delta/2}\over{\Delta/2}}.\eqno(7)
$$
$$
E={{ea}\over{\hbar}}E_{sc}+\omega_{s}\cos\Omega t,\quad \omega_s={{dE_0}\over{\hbar}}, \quad d=ea.\eqno(8)
$$
In these variables, Eqs. (3) become
$$
\dot v=Ew-\gamma_vv,\eqno(9a)
$$
$$
\dot w=-Ev-\gamma_\varepsilon(w-w_0),\eqno(9b)
$$
$$
\dot E=\omega^2_Ev-\alpha E+f(t),\eqno(9c)
$$
where
$$
f(t)=\alpha\omega_s\cos\Omega t-\omega_s\Omega\sin\Omega t \quad {\rm and}
\quad
\omega_E=\left[{{2\pi e^2Na^2\Delta}\over{\hbar^2}}\right]^{1/2}.\eqno(10)
$$
It follows from Eq. (5) that $w^{(T)}_0=2\varepsilon^{(T)}_0/\Delta-1$.
The variable $w$ in Eq. (7) has a simple physical interpretation: namely, it is the
electron's energy measured from the middle of the miniband and normalized
by the half-width of the miniband. Hence the value $w=-1$ corresponds to the bottom
of the miniband, and $w=1$ corresponds to the upper edge of the miniband.
The field $E$ is the total electric field (measured in units of frequency)
acting on the electrons. Note that the frequency $\omega_E$ is {\it formally}
equal to the
frequency of electron {\it plasma oscillations},  $\omega_{pl}=[4\pi e^2N/m_0]^{1/2}$,
{\it provided} that
$m_0$ is taken as the effective electron mass at the bottom of the miniband,
as given by Eq. (4). For this reason, the corresponding cooperative oscillations
of the coupled SSL system were called ``nonlinear plasma
oscillations'' by Epshtein \cite{epstein2}. However, since the term
``plasma oscillations'' is usually used in a different context
in semiconductors \cite{ferry1,vatova} we shall refer to these as
``cooperative oscillations'' and denote their frequency by $\omega_E$.

The analogy mentioned in the introduction between the present SSL
problem and optical systems helps to clarify this point further.
If we consider the variable $w$ in
Eqs. (9) to be the population difference and $v$ the polarization, then
Eqs. (9)
are equivalent to the coupled Maxwell-Bloch (CMB) system, taking into account the
external field (see, for example, \cite{gibbs}).
Recall that the CMB equations describe the dynamics of two-level
atoms placed in a single-mode cavity and interacting with the cavity field
via a dipole interaction.
The width of the SSL miniband ($\Delta/\hbar$) is equivalent to the transition frequency
($\omega_0$) of the two-level atom. The value $ea$ of the SSL is the transition
dipole moment ($d$) in the CMB case. Finally, the analog of the
frequency $\omega_E$ defined by Eq. (10) is the so-called ``cooperative frequency'',
$\omega_c=(2\pi\omega_0d^2N/\hbar^2)^{1/2}$ \cite{dicke}.
The physical interpretation of this cooperative frequency is
the frequency of the slow resonant exchange of energy between the $N$ two-level
atoms and the field in the cavity, as first described by the Dicke model \cite{dicke}.
In the balance equations describing
the SSL system (Eqs. (9)), the miniband is treated as initially empty.
When the $N$ electrons are injected, the assumption that they are distributed
with spatially homogeneous density means that the ``populated'' miniband
becomes analogous to an optical system consisting of an ensemble
of two-level atoms with density $N$ in the Dicke model. Our self-consistent
field is analogous to the cavity field, and the cooperative
oscillations of the Dicke type appear due to the coupling
of the electrons in the miniband to the self-consistent field. The
relaxation parameter $\alpha$ in our SSL equations plays the
role of the finite quality factor of the cavity in the optical system.
Finally, the Stark frequency $\omega_s=eaE_0/\hbar$ is
equivalent to the Rabi frequency $\omega_R=dE_0/\hbar$.

Eqs. (9) are also similar to the CMB equations describing the dynamics of
a laser with an injected signal \cite{abraham},
or a bistable system in the
framework of the Bonifacio-Lugiato model \cite{gibbs,benza}.
However, there are important differences between Eqs. (9) and the optical
analog systems. First, the CMB equations are derived for the field and the
polarization envelopes; hence, the variables $v$ and $E$ in the
CMB equations are generally complex. Second, the form of the external force
$f(t)$ that perturbs the cavity mode is different in the CMB equations.
Nonetheless, the structural similarities and the
well-known results that under certain conditions
transitions to chaos take place in the bistable devices \cite{gibbs,benza}
and also in lasers both with \cite{lugiato} and
without injected signals \cite{haken1,haken2}
suggest that one should expect transitions to chaotic dynamics in
the SSL plus field interaction problem.  In the next section, we will
show that this is indeed the case.

\section{CHAOTIC DYNAMICS IN SEMICONDUCTOR SUPERLATTICES}

Exploring the chaotic dynamics in Eqs. (9) is a formidable task,
for we have in effect a four-dimensional dynamical system -- three
independent variables ($v,w,$ and $E$) plus the explicit
external time-dependence --
involving six parameters ($\gamma_v, \gamma_{\varepsilon},\alpha,
\omega_E, \omega_s$, and $\Omega$). Fortunately, we can obtain
some guidance concerning the ``interesting'' regions of parameter and
state space by noting that our equations reduce, in two different
limits, to two well-known systems which exhibit chaos. We discuss
these two limits in the ensuing two subsections.

\subsection{The Lorenz Equation Limit}

When the external ac field is absent ($f(t)\equiv 0$),
Eqs. (9) are equivalent to the well-known Lorenz model \cite{lorenz}.
Although one can derive this equivalence directly, it can also be seen immediately
from our optical analogy: namely, for $f(t)\equiv 0$, Eqs. (9) coincide
with the CMB equations describing a single-mode homogeneously broadened laser at
exact resonance between the cavity mode and two-level atomic transition \cite{haken1},
and it has been shown by Haken \cite{haken2} that the CMB equations can be
reduced to the Lorenz model by a simple transformation of variables.

Translating the necessary conditions for instability and the transition to
chaos in the Lorenz model \cite{haken1,haken2,lorenz}
into our notation, we find that these conditions can be written as
$$
\alpha>\gamma_v+\gamma_\varepsilon,\eqno(11a)
$$
\vspace{0.2in}
\noindent
and
$$
w_0>{{\gamma_v \alpha^2}\over{\omega^2_E}}{{(\alpha/\gamma_v+\gamma_\varepsilon/\gamma_v+3)}\over{(\alpha-\gamma_\varepsilon-\gamma_v})}.\eqno(11b)
$$
The condition in Eq. (11b) requires that the value $w_0$ corresponding to the
electron's equilibrium energy should be larger then some critical value,
$w^{(cr)}_0>0$. From (5), we see that even at high temperatures
($T\rightarrow\infty$), the equilibrium value $w_0^{(T)}\rightarrow -0$.
Hence we find the interesting result that in the Lorenz limit ($f(t) \equiv 0$),
the necessary conditions (11b) for the transition to chaos can not be satisfied
in the SSL system; to obtain chaos in
the SSL system, we require additional driving ($f(t) \neq 0$).

\subsection{The Damped, Driven Josephson Junction Limit}

When $\gamma_v=\gamma_\varepsilon=0$, the Eqs. (8a) and (8b) immediately
imply the existence of a constant of motion: namely, the length
of the pseudo-spin vector, which we can without loss of
generality scale to 1, so that
$$
v^2(t)+w^2(t)=1.\eqno(12)
$$
We can incorporate this conservation law explicitly and consistently into the
dynamics by introducing the change of variables
$$
v=-\sin\theta,\quad w=-\cos\theta,\quad \theta=\int_0^tdt^\prime E(t^\prime).\eqno(13)
$$
From these definitions (and noting that $\dot\theta=E$),
one sees that Eqs. (9a) and (9b) are automatically satisfied and that
Eq. (9c) becomes
$$
\ddot\theta+\alpha\dot\theta+\omega_E^2\sin\theta= \omega_s(\alpha\cos\Omega t-
\Omega\sin\Omega t).\eqno(14)
$$
Introducing $\phi_0= \tan^{-1}(\Omega/\alpha)$, we can recast Eq. (14)
into the form
$$
\ddot\theta+\alpha\dot\theta+\omega_E^2\sin\theta= \rho \cos(\Omega t+ \phi_0), \eqno(15)
$$
where $\rho = \omega_s\sqrt{\alpha^2+\Omega^2}$. This
is the canonical form of the damped, (ac) driven pendulum equation, widely
studied in chaotic dynamics both in its own right and as a model for
a damped, driven Josephson junction \cite{huberman}. We shall exploit this
connection further in our detailed analysis of the numerical results below.

In the limit of no dissipation whatever -- so $\alpha = 0$ as
well as $\gamma_v$ and $\gamma_{\varepsilon}$ --- the SSL plus field
system reduces to the Hamiltonian system which we studied in
Ref. \cite{alekseev1} for a range of physically reasonable
initial conditions; interested readers should consult this
reference for details.
Here we simply remark that this non-dissipative chaos in the SSL plus field
system has its own optical analogy, involving the generalized
semi-classical Tavis-Cummings (TC) model, which describes the dynamics of two-level
atoms in a single-mode high quality cavity, interacting with self-consistent and
external fields \cite{alekseev2}.
Details of the chaotic dynamics in the TC model can be found in
\cite{alekseev2,alekseev3}.

It is well known that both Eq. (15) and its undamped counterpart
contain chaotic dynamics \cite{chirikov}. In terms of
our parameters, the parameter region in which strong chaos is
expected is $\omega_E\sim\omega_s\sim\Omega$; in the next subsection,
we shall use this information as the starting point for our
study of chaos in the full SSL plus field problem.

\subsection {Dissipative Chaos in the Presence of an External Time-Periodic Field}

We now consider the general case, in which all of
$f(t),\gamma_v,\gamma_\varepsilon$, and $\alpha $ are non-zero. The
structure of the Eqs. (9) suggests that we take $\omega_E$ as the
scale of (inverse) time, and thus the natural damping parameters
that occur in the re-scaled equations are the dimensionless
quantities $\gamma_v/\omega_E, \gamma_\varepsilon/\omega_E $
and  $\alpha/\omega_E$. Since there is considerable uncertainty
in the individual values of the phenomenological damping parameters,
we will study a broad range of values
of these parameters: $ 0 \leq \gamma_v/\omega_E, \gamma_\varepsilon/\omega_E
\leq 0.2$ and $ 0 \leq \alpha/\omega_E \leq 0.2$.
Since there are as yet no direct measurements of the cooperative
oscillations or their damping, we have chosen conservative upper damping
limits inferred from recent results
determining that the ratio of the line width of the
{\it plasma} oscillations to their frequency
can be as large as  $ 2 \times 10^{-1}$ \cite{quinn,camley}.
For initial conditions, we take $E(0)=\omega_s, v(0)=0, $ and
$ w(0)= -1$, corresponding to the initially unexcited SSL just
being struck by the incident EM radiation. For our numerics
we used a fifth-order Runge-Kutta algorithm incorporating
adaptive step size, accuracy checking, and Cash-Karp optimized
parameters.

The variable most directly related to experimental observables
is the average electron velocity, $v$. Accordingly, we will focus on the various
different behaviors of $v$ that follow from the solutions
of Eqs. (9) and the regions in which
they occur. In Fig. 1 we show the two basic types of
behavior for $v$ observed in our simulations. Fig. 1a shows
behavior in the ``regular'' region, in which
the velocity varies periodically. In Fig 1a, the basic
frequency is just the fundamental frequency of
the external EM field
with a longer period modulation (caused by the nonlinearity
of the equations) superimposed. The ``locking'' of the oscillations of the
electron's velocity to the fundamental frequency of the external field
is referred to as ``1:1 mode-locking behavior'' in the Josephson
junction literature \cite{huberman} and in related studies of coupled
oscillators. Fig. 1b shows a typical behavior in the ``chaotic'' region,
in which the velocity varies erratically and with
no apparent periodicity for as long as we observe it; this
is ``stationary chaos'' and is the behavior in which we are most interested.
Within the region of parameters in which chaos is observed,
we also observe a behavior which exhibits characteristics
of both the regular and the stationary chaotic motion: namely,
a (typically long) interval of ``erratic'', aperiodic motion, followed
by a near
vanishing of the oscillations and then a locking into
a periodic motion; this {\it transient chaos} is illustrated in Fig. 2.
Importantly, the time at which the transient chaos disappears, $t_{tr}$ is
a sensitive function of the numerics, especially the level
of accuracy demanded of the numerical integrator. This is
commonly encountered in simulations of chaotic systems.

In the regions of transient chaos, the asymptotic state is periodic.
Visual inspection of Fig. 2b shows that for these parameter values the final
period is also the fundamental of the external period, but
we have also observed (for other parameter values)
locking to different subharmonics of this period. The general case
of locking into periods other than the fundamental period of the
external drive is well-known from general results in nonlinear
dynamics and is exhibited explicitly by the damped, driven
Josephson junction \cite{huberman}. Given our present focus
on establishing the possibility of chaotic motion in the SSL
system, we shall not present further details here \cite{cannon}.

To quantify these three types of behavior systematically,
we used standard dynamical systems tests: for each set
of parameters, we calculated
the maximum Lyapunov exponent, $\lambda$, using the method
described in Ref. \cite{wolf} and determined the power spectrum
(using an FFT algorithm) for each of the velocity plots.
In Fig. 3 we show the typical behavior
of the maximal Lyapunov exponent for the cases of regular motion and
chaotic motion. With the standard definitions and calculational
procedures \cite{wolf}, the Lyapunov exponent will vary in
time, eventually converging to the value reflecting the underlying
long-time dynamics. We see from Fig. 3
that for the chaotic motion, the asymptotic
value of $\lambda$ is greater than zero, as it should be, whereas
for the periodic behavior it is less than zero. For the parameter
values chosen in Fig. 3a, this asymptotic value is only slightly
negative, consistent with the fairly weak dissipation for these
values of the parameters.
For the case of transient chaos, shown in Fig. 3b, the Lyapunov
exponent decays to its final value only very slowly.

In Fig. 4 we show typical power spectra for periodic and
chaotic motion.
Note the expected appearance of a broad power spectrum in the
chaotic case, in contrast to the isolated peaks associated
with the periodic evolution.

The best overview of the qualitative nature -- chaotic versus
periodic -- of the behavior of the system is provided by a
two-dimensional plot showing the locations of the regions with
{\it positive} values of the Lyapunov exponent as functions of the two
parameters of the external field, $\omega_s$ and
$\Omega$, measured in units of $\omega_E$. This sort
of ``$\lambda$ topographic map'' -- henceforth, ``$\lambda$-map'' --
has been used very effectively
in studies of chaos in the damped, driven Josephson junction \cite{huberman}. 
Here it will allow us readily to see how various types
and amounts of damping effect the extent of chaos in our system.
Further, viewing the $\lambda$-map as a topographic map
suggests a colorful but informative
terminology in which the region of the map with $\lambda > 0$ is
viewed as a ``continent,'' with the surrounding ``sea'' being
the region with $\lambda < 0$, in which the velocity behaves as
a periodic function of time with the same period as the
external drive. On the ``dry land'' of the continent,
the velocity exhibits stationary chaos, but
within the continent, there are (depending
on parameters) ``geographical'' features -- such as fjords,
deltas, lakes, and channels -- in which the velocity again (typically after
an interval of transient chaos) shows periodic behavior,
in some instances with a subharmonic of the external driving
frequency. 

In Fig. 5 we present the first of the $\lambda$-maps for our system.
For purposes of comparison with Ref. \cite{huberman},
we have chosen the parameters to correspond
to the case of the Josephson junction ($\gamma_v=\gamma_{\varepsilon}=0$)
and have produced a $\lambda$-plot in the ($\rho - \Omega$) plane instead of the
($\omega_s - \Omega$) plane; again, the plots are in units of $\omega_E$.
As stressed above, this is not a physically plausible set of parameters for real a
SSL because it neglects crucial velocity and energy dissipation
effects, but it does provide a convincing test
of our numerics, for direct comparison shows
that our results (plotted here with $\alpha/\omega_E =0.2$) are
in full agreement with those of Ref. \cite{huberman}
as far as concerns the structure of the chaotic
regions; further, although we shall not present the details of
the harmonics in the periodic
regions because they are not germane to our current discussion,
we have also found \cite{cannon} good agreement with Ref. \cite{huberman} for the
periodic regions. Qualitatively, Fig. 5 shows us the triangularly
shaped southern extremity of a large chaotic continent (which extends upwards
for toward larger values of $\rho$). The southwest
edge of the continent looks like a river delta, with many periodic
channels cutting through the chaotic region; these channels correspond
to the different subharmonic periodic lockings observed in
simulations of Josephson junctions.

Turning to parameter values more relevant to SSLs, we
show in Fig. 6a-d the evolution of the chaotic region
as the values of the damping parameters are 
varied over a wide range. In Fig. 6a
we begin from the limit of fairly small dissipation
($\gamma_v=\gamma_{\varepsilon}=0.01$, $\alpha= 0.001$). Note
that in Fig. 6 and henceforth, all damping parameters are
scaled in units of $\omega_E$.
In this case, the ``order-chaos'' boundary
is very close to the boundary found in \cite{alekseev1}
for the Hamiltonian model; given the fairly small values of the relaxation
parameters, this is perhaps not terribly surprising.
Further, the chaotic continent is very solid, with few of the
channels observed in Fig. 5 (or in subsequent figures discussed below).

In Fig. 6b all relaxation rates have been increased by
a factor of 10 (so that $\gamma_v=\gamma_{\varepsilon}=0.1$, $\alpha= 0.01$).
The $\lambda$-map shows substantially more structure in the
chaotic regime, with a number of fjords cutting into the
west coast, and the east coast receding substantially, leaving
effectively an offshore island of chaos. Increasing the damping
still more reduces the chaotic region, first to the narrow peninsula
plus islands in Fig. 6c and then to the two tiny islands of chaos
in Fig. 6d. These figures make clear that while the size and shape of
the chaotic region are strong functions of damping, chaos
is expected to occur for a wide range of values. In particular,
for the expected range of the velocity and energy damping
parameters, $ 0.01 \leq \gamma_v, \gamma_{\varepsilon} \leq 0.1 $
\cite{quinn,camley} found in the highest quality SSLs, chaos appears
likely to occur over a substantial range of parameters.

Let us comment on several qualitative features of the chaotic regions in
Fig. 6, beginning with Fig. 6a. Perhaps the most striking qualitative
feature here is the clear distinction between the ``west coast'' of
the chaotic continent, which appears sharply defined (like a set of cliffs) , and the
``east coast,'' which is substantially more diffuse (like a river delta).

An enlargement of the east coast region, shown in Fig. 7a, suggests that
the river delta may have a fractal boundary, since zooming in on
the region does not decrease its structure. Although this structure
is theoretically interesting, the experimental consequences are likely to be 
limited. First, in this region, many of the positive Lyapunov exponents
are nearly zero and are thus sensitive to small effects from
the numerics. It is thus difficult to be certain of the boundaries between periodic and
chaotic behavior. To illustrate the sensitivity to a cut-off on the
size of $\lambda$, we show in Fig. 7b the same enlargement of the
east coast with the constraint that  $\lambda > 0.01$. The difference
between Fig. 7a and 7b is readily apparent. Second, since,
as shown in Fig. 3b and discussed above,
the Lyapunov exponents can sometimes relax very slowly to their
asymptotic values -- leading to regions of ``transient chaos''--
determining the true asymptotic value of  $\lambda$ can be difficult.
Indeed, the precise boundary in Fig. 7a is very sensitive to the details
of the numerical code, including discretization effects, and
a different code would likely not reproduce it exactly. Particularly
if there is a fractal boundary \cite{grebogi} for the actual differential
equation system, this effect is to be expected.

In contrast, the west coast in Fig. 6a is very sharply defined:
the negative exponents jump suddenly
to large positive ones as the coastline is crossed.

Our interpretation of these features can be described qualitatively
in terms of familiar concepts from dynamical systems. Recall
that in exhaustive studies of general nonlinear dissipative
dynamical systems one fixes the parameters and varies the initial conditions, searching
for all the ``attractors'' and determining the shape of each basin
of attraction in the space of initial conditions. Typically there
is more than one attractor in the system, and the boundaries between
the different basins of attraction can be smooth or fractal \cite{grebogi}. After
determining completely the attractor structure for one set of parameters,
one then moves on to another set, and does a similar search through
the space of initial conditions.
In our study, which is intended to be illustrative of the possible
existence of chaos in SSLs (rather than exhaustive),
we have for simplicity {\it fixed} the initial conditions
and varied the parameters. As a consequence, if there are multiple
attractors, as we change parameters for fixed initial conditions,
we can pass from one basin of attraction another, and the structure
of the basin boundaries --fractal or smooth-- will be reflected in the $\lambda$-maps.
More extensive numerical studies, which will shall report elsewhere \cite{cannon},
confirm this explanation.

In Figs. 6b-c, the size of the chaotic continent decreases successively,
and fjords become apparent. These figures -- particularly Fig. 6b --
interpolate nicely between the case of the damped, driven Josephson junction (Fig. 5)
and the case of Fig. 6a. The presence of these fjords raises
the possibility of observing not only chaos in SSLs but also
mode-lockings to various subharmonics
of the driving frequency, and we are currently investigating this possibility
\cite{cannon}.

In real SSLs, one typically has $\gamma_v >> \gamma_{\varepsilon}$. In our
previous discussion, we have for simplicity considered only the
case $\gamma_v = \gamma_{\varepsilon}$. Figure 8
illustrates the extent of the chaotic region when
$\gamma_v = 0.1 >> \gamma_{\varepsilon} = 0.02 $; this reduction
in the chaotic region for unequal damping is typical \cite{cannon}. 
Restricting ourselves to the physically relevant regime in which
$\gamma_v > \gamma_{\varepsilon}$, we can summarize our data qualitatively by saying
the region of chaos is largest when $\gamma_{\varepsilon}$ is roughly equal to
$\gamma_v$ or $\alpha$, whichever is larger.

Finally, we note that for simplicity in all the above results
we worked at at zero temperature.
From Eq. (5), we see that this corresponds to $\varepsilon_0=0$ ($w_0=-1$).
For $\Omega/\omega_E=1$ and for various values of $\omega_s/\omega_E$, we
investigated the influence of the temperature effects on the
transition to chaos. When $T$ was varied from helium to the room temperature,
keeping all other parameters fixed, we found no qualitative changes
in the nonlinear dynamics within our phenomenological balance equation model.
Quantitative details of the temperature dependence will be presented
elsewhere \cite{cannon}.

\section{SUMMARY, DISCUSSION, AND CONCLUSION}

We have considered the influence of an ac electric field on the motion of
ballistic electrons in a miniband of a semiconductor superlattice.
Within a phenomenological balance equation approach,
we established that accounting for collective effects (via a self-consistent
field) leads to the possibility of chaotic dynamics. Our numerical and
analytic results suggest that for a transition to chaos one
must satisfy the following conditions: (i) the frequency of the ac field ($\Omega$) should
be close to the characteristic frequency of the collective electron motion
($\omega_E$) in the
SSL; (ii) at the same time, the
frequency of the ac field should be close to the Stark frequency
$\omega_s=eaE_0/\hbar$, which is determined by the {\it amplitude} of the external
field; and (iii) the relaxation rates of the electron's energy and momentum should not
be too large ($\gamma/\omega_E \stackrel{<}{\sim}0.2$).

Importantly, it appears possible to achieve these conditions in real SSLs, now
or in the near future. For typical superlattices ($a\sim 10^{-6}cm$, $\Delta\sim 10^{-2}eV$, $N\sim 10^{14}cm^{-3}$),
the characteristic frequency of the cooperative oscillations lies in the THz
domain ($\omega_E\approx 1.5\times 10^{12}sec^{-1}$)
\cite{epstein1,quinn,perkowitz}.
If an ac field with amplitude
$\sim 1kV/cm$ is applied to an SSL with period $a\sim 10^{-6}cm$,
the Stark frequency also lies in the THz domain. Thus the frequency
constraints can likely be achieved.

Although the damping effects and relaxation rates are much less
well known, there nonetheless appears to be reason for some optimism.
As we have indicated above, standard
estimates of the relevant relaxation constants for the plasma oscillations
give values in the range of $ (10^{-1} - 10^{-2})\omega_{pl}$
\cite{quinn,camley}. That the damping is more likely near the
larger end of this range is suggested by the observation that
the phase relaxation
rate even in a good quantum well is the relatively
rapid $\tau_v \simeq 3.5 \times 10^{-12}$ seconds
\cite{sherwin}, which corresponds to near THz frequencies. For
a modulation-doped superlattice, $\tau_v$ is almost certain to be shorter,
since the electrons may scatter from dopant impurities not present in
the remotely doped quantum wells. Hence, the damping effects
in current SSLs may be nearer the high end of our range of
parameters. One intriguing possibility for producing SSLs with
lower relaxation rates involves ``implanting'' the
superlattices within parabolic quantum wells
\cite{campman}; in this manner, one could hope to achieve the low
damping levels of good quantum wells and avoid
the damping effects associated with modulational
doping. A separate complication concerning
damping effects is that energy relaxation
processes in both wells and superlattices
also involve many distinct processes. From quantum well studies \cite{sherwin},
we expect that the energy relaxation can not be
described {\it  quantitatively} by a single, constant $\gamma_{\epsilon}$.
Although it would clearly be possible to introduce energy-dependent
relaxation rates into our phenomenological equations, without
more detailed experimental guidance as to the form of this
dependence, it is premature to incorporate
such an additional complication. Given the relatively large range of
relaxation parameters over which our phenomenological model predicts chaotic
behavior for the SSL plus field system, the present uncertainties
in the exact level and nature of damping in these systems are not
cause for undue concern.

Thus, we believe that by applying
an ac field of order $\sim 1kV/cm$ with frequency of the order
of several THz to an SSL, one should
be able to satisfy the requirements for transition to deterministic chaos.
For instance, in the recent experiment \cite{ignatov1} first
studying the influence of a THz-field on the stationary electron transport
properties
in an SSL, the experimental conditions were close to those
required for the observation of chaos in our model system.
Further, continuing progress in both the fabrication of heterostructures with
high carrier mobility and in the design of powerful sources of
THz-radiation \cite{sherwin}
suggests that the experimental observation of the deterministic
chaos in SSL plus field interaction may be close at hand.

An essential question for experimentalists is how to recognize the
underlying chaotic electron dynamics in the observables
measured in a real experiment. As the controversy
over the observation of Bloch oscillations suggests, this may not be
a simple matter. In large part, it will depend on precisely
how the experiment is configured and instrumented. At this stage,
and without considering in
detail the configuration of a particular proposed experiment, we can
most appropriately give a somewhat general answer. If chaos is
present, we expect a complex, aperiodic behavior for
the average velocity and hence the average current. Given the high
frequencies involved, it seems unlikely that one could
measure this directly in the time domain. However, in the presence
of an additional dc voltage, to create non-zero mean ``drift'' in this average 
velocity, the oscillatory chaotic component would appear
as a substantial additional source of apparent ``noise'' in
drift velocity and that this additional noise would appear
suddenly as one crossed the threshold to chaos, particularly
for the parameter regime corresponding to the ``west coast'' of
the chaotic ``continent.''
Specifically, if one measured the power spectrum associated with
the current, one would observe the same substantial increase in
the broad-band ``noise'' component that we see in Fig. 4.
In an experiment on a resistively shunted Josephson tunnel
junction (related to the model in our Eq. 15) precisely such a
dramatic increase in
experimental noise was observed when the paramters of the experimental
system were moved through the transition to chaos \cite{miracky}.
A second option for detecting chaos would involve
sampling the current at given time intervals and using the
``phase space reconstruction'' techniques \cite{packard} to create
a geometrical image of the underlying attractor. For regular motion,
the attractor will be a simple periodic structure, for chaotic
motion, it will be a ``strange attractor.'' The details of this
approach are described in the context of an experiment
involving germanium photoconductors in Ref. \cite{westervelt}
Additional details about experimental techniques for detecting
chaotic motion in semiconductor structures are described in
Ref. \cite{sherwin}.

Apart from the most central issue of experimental verification
of the existence of chaos, there are a number of open theoretical
issues which merit further study. First, our model is applicable
in the limit of {\it miniband} transport for the electrons
and assumes a spatially homogeneous structure for the EM field.
One could ask whether these assumptions are crucial to the
possible existence of chaos in SSLs. A recent study by
Bulashenko and Bonilla \cite{bulashenko} suggests
strongly that this is not the case. Focusing on the resonant-tunneling
regime and considering the possibility of high-field
domain effects (non-homogeneity in space), these authors
also find possible chaotic behavior.
Since the cross-over between the resonant-tunneling and miniband regimes
of transport depends on many experimental and material parameters,
some more controllable than others, and remains a complicated
problem for theorists, it is important to note that these
essentially complementary studies,
taken together, suggest that chaos is a robust phenomenon in SSLs.
Second, there is a clear need to understand the extent to
which our phenomenological balance equation approach
correct captures the physics contained in more microscopic
considerations, such as the full Boltzmann equation or an approach
based on Wigner distribution function \cite{ferry2}, and on
the anticipated region of validity for any miniband-based
approach. Although there has been some recent progress on the
former issue \cite{lei2}, there remains much to be done on both
problems.

Let us conclude with a brief speculative comment
related to the possible consequences and relevance of chaotic
behavior in SSLs. Based on the earlier experience of studying
chaos in semiconductor devices used for infra-red
radiation detection \cite{westervelt},
mapping out the boundary of chaos experimentally is important to
reliable use of the devices in the ``normal'' regime. However, 
recent developments in ``controlling chaos'' \cite{shinbrot} suggest
that one might actually deliberately choose to drive the SSL
into a chaotic regime, in order to take advantage of the myriad
possible behaviors there for device applications. Alternatively,
using methods of chaotic control, one may be able to suppress the
onset of chaos, as was recently done in an experimental laser
system \cite{roy}.
Chaotic control and reduction of chaos are also likely
to be important for future nano-fabricated semiconductor integrated circuits, 
where the expected chip densities will 
be of the order $10^9/cm^2$ \cite{ferry1}. At such densities,
the devices actually form a {\it lateral surface superlattices} (LSSL),
and device-device interactions can generate both cooperative effects
and additional instabilities \cite{ferry7}.
In any case, a large number of exciting
experimental and device-related problems remain.

\section{ACKNOWLEDGEMENTS}
It is a pleasure to thank Mark Sherwin for many valuable discussions
on experimental aspects of SSL's. We also gratefully acknowledge
useful comments from Feo Kusmarsev, Jesper Myqind, Mads Sorensen, and
Stephanos Venakides. KNA and GPB thank The Center for Nonlinear Studies, Los
Alamos National Laboratory, and Department of Physics
at The University of Illinois at Urbana-Champaign, for their hospitality. This
work was partially supported by the
Grant 94-02-04410 of the Russian Fund for Basic Research and by
the Linkage Grant 93-1602 from the NATO Special Programme Panel on
Nanotechnology. EHC thanks the US Department of Education for
support by a GAANN Fellowship (DE-P200A40532), and MCC thanks
the US-NSF for support under its REU program (NSF PHYS93-22320).

\newpage

{\bf Figure Captions}\\ \ \\
Fig. 1. Dependence of the electron's average velocity on time for
$E(0)=\omega_s$, $v(0)=0$, $w(0)=-1$; (a) ``regular'' (periodic) dynamics
(for $\omega_s/\omega_E=0.1$, $\Omega/\omega_E=1$);
(b) chaotic dynamics (for $\omega_s/\omega_E=1.6$, $\Omega/\omega_E=0.2$).\\ \ \\
Fig. 2. Dependence of the electron's average velocity $v$ on time for
transient chaos ($\gamma_v/\omega_E= \gamma_{\varepsilon}v/\omega_E = 0.01$,
$\alpha/\omega_E=10^{-3}$, $\omega_s/\omega_E=1.5$, $\Omega/\omega_E=1$);
(a) long-time behavior;
(b) transition to the laminar phase.\\ \ \\
Fig. 3. Dependence of the maximal Lyapunov exponent on time; (a) for
chaotic motion (solid curve; the parameters are the same as in Fig. 1b); and
for periodic motion (dashed curve; for $\omega_s/\omega_E=1.3$,
$\Omega/\omega_E=0.2$); (b) for the transient chaos shown in
Fig. 2.\\ \ \\
Fig. 4. Power spectrum versus frequency for the electron's velocity $v$; (a) Regular
motion (parameters are the same as in Fig. 1a); (b) Transient chaos (parameters
are the same as in Fig. 2). For the case of the transient chaos, the frequency
spectrum was calculated only for the turbulent phase.\\ \ \\
Fig. 5. The $\lambda$-map showing the regions of periodic (white) and
chaotic (symbols) motion in the $\rho$-$\Omega$ plane; the values of
the damping constants are  $\gamma_v = \gamma_{\varepsilon} = 0$ and
$\alpha = 0.2$, corresponding to the case of the damped, driven
Josephson junction studied in Ref. \cite{huberman}. In this and
all subsequent figures, all parameters are measured in units of $\omega_E$. \\ \ \\
Fig. 6. $\lambda$-maps showing the regions of periodic (white) and
chaotic (symbols) motion in the $\omega_s$-$\Omega$ plane for
four different values of the damping parameters:
(a)  $\gamma_v = \gamma_{\varepsilon} = 0.01, \alpha = 0.001$;
(b)  $\gamma_v = \gamma_{\varepsilon} = 0.1, \alpha = 0.01$;
(c)  $\gamma_v = \gamma_{\varepsilon} = 0.1, \alpha = 0.05$; and
(d)  $\gamma_v = \gamma_{\varepsilon} = 0.2, \alpha = 0$; \\ \ \\
Fig. 7.  $\lambda$-maps showing an enlargement of the ``east coast'' 
of the chaotic region for the parameters
$\gamma_v = \gamma_{\varepsilon} = 0.05, \alpha = 0.01$ as determined
by requiring (a) $\lambda > 0.001$; and (b) $\lambda> 0.01$. \\ \ \\
Fig. 8.  $\lambda$-maps showing the chaotic region in the case
in which the elastic scattering time is considerably larger than
the inelastic scattering time. The parameters are 
$\gamma_v = 0.1, \gamma_{\varepsilon} = 0.02, \alpha = 0.01$. \\ \ \\

\end{document}